\documentclass[11pt,a4paper]{article}
\usepackage[dvips]{graphicx}
\title{
Control of photodissociation branching 
using the complete reflection phenomenon: 
Application to HI molecule
} 
\author{Hiroshi Fujisaki$^a$, 
Yoshiaki Teranishi$^b$,
and Hiroki Nakamura$^{a,c}$
\\
$^a${\it Department of Theoretical Studies,
Institute for Molecular Science},
\\
{\it Myodaiji, Okazaki 444-8585, Japan}
\\
$^b${\it Advanced Photon Research Center, 
Japan Atomic Energy Research Institute},
\\
{\it Kizu-cho, 
Kyoto 619-0215, Japan}
\\
$^c${\it Department of Functional Molecular Science},
\\
{\it The Graduate University for Advanced Studies,
Myodaiji, Okazaki 444-8585, Japan}
}
\begin{document}
\maketitle

\begin{abstract}



The laser control of photodissociation branching in 
a diatomic molecule is demonstrated to be 
effectively achieved with use of the complete 
reflection phenomenon. The phenomenon and the control
condition can be nicely formulated by the semiclassical
(Zhu-Nakamura) theory. The method is applied to the 
branching between 
I($^2 P_{3/2}$) (HI $\rightarrow$ H + I) and 
I$^*$($^2 P_{1/2}$) (HI $\rightarrow$ H + I$^*$)
formation, and nearly complete 
control is shown to be possible 
by appropriately choosing an initial vibrational state
and laser frequency in spite of the fact
that there are three electronically excited states involved.
Numerical calculations of the corresponding 
wavepacket dynamics confirm the results.

\end{abstract}

\section{Introduction}


Controlling chemical dynamics \cite{RZ00}
is one of the hot topics recently in chemical physics.
Since the works of Brumer and Shapiro 
(coherent control) and Tannor and Rice
(pump-dump method), a lot of 
studies have been made on the subject both theoretically and 
experimentally.
The essence of control consists in transfering 
a certain 
prepared (initial) quantum state into a desired (target) quantum state, 
in which the coherent interaction between a molecule and 
a laser field plays a crucial role.
Each one of the ideas proposed so far 
has its own advantages and disadvantages.
For instance, the optimal control theory initiated by 
Rabitz, Kosloff, Rice and others can, {\it in principle}, 
deal with any kind of target, 
but its numerical cost becomes formidable
for more than 3D systems, and the optimal field 
generated is often a bit too complicated to be realized in 
experiment. 
Our basic idea, on the other hand,
is to regard the molecular processes in a laser field
as a sequence of nonadiabatic transitions 
and to control them.
This concept can be justified by the Floquet theory
for periodically perturbed systems, and introducing 
the Floquet (or dressed) states \cite{TN99,NTN00,TNN01,NTN02a,NTN02b,ZTN01}.  
Namely, we try to control nonadiabatic transitions among dressed states. 

Previously we have proposed such methods of control 
with the help of the semiclassical theories of 
nonadiabatic transitions 
including the Zhu-Nakamura theory as a complete solution of 
the Landau-Zener-St\"uckelberg (LZS) type 
curve crossing problems \cite{ZTN01,Nakamura02}. 
They are categorized as 
the method using (a) the time-dependent and 
(b) the time-independent 
theory of nonadiabatic transitions.
In the former case \cite{TN99,TNN01,NTN02a,NTN02b,ZTN01},
periodical sweeping of laser parameters or a sequence   
of linearly chirped pulses are employed to control
various processes.
In the latter case \cite{NTN00}, 
the complete reflection phenomenon in the time-independent 
nonadiabatic tunneling type transition \cite{ZTN01,Nakamura02}
has been employed 
to discuss the possibility of controlling 
photodissociation by using the {\it model} 1D and 2D 
molecules, i.e., models of HOD and CH$_3$SH.
However, up to now, there has been no application of the theory 
to a real molecule.
In this paper, following the same strategy 
as in \cite{NTN00}, we try to control 
photodissociation branching 
of a real HI molecule with use of the 
complete reflection phenomenon and 
ab initio data of the potential curves.
The complete reflection phenomenon, the existence of
which was quantum mechanically exactly proved by Zhu and 
Nakamura \cite{ZN92}, can be accurately formulated by the 
Zhu-Nakamura semiclassical theory \cite{ZTN01,Nakamura02} 
(see also \cite{Nakamura87,Ovchinnikova65,BC78}) and has been utilized 
to propose a new molecular switching \cite{Nakamura92,NNG97,Nakamura99}.

This paper is organized as follows:
In Sec.\ 2, we briefly summarize the semiclassical theory 
and the phenomenon of complete reflection.
In Sec.\ 3, we describe an HI molecule in a laser field where 
the ab initio 
data for HI calculated by Alekseyev {\it et al}.\ \cite{ALKB00} 
are also shown.
With use of the Zhu-Nakamura theory,
we find effective conditions for 
controlling photodissociation branching:
HI $\rightarrow$ H + I/HI $\rightarrow$ H + I$^*$. 
In Sec.\ 4, we report the results of 
full numerical solutions of wave packet dynamics.
The semiclassical prediction is confirmed to be accurate,
and the present control scheme is proved to be effective.
Section 5 is devoted to concluding remarks and discussions.

\section{Complete reflection in a diatomic molecule 
--- Semiclassical theory of nonadiabatic transition}

\begin{figure}[htbp]
\hfill
\begin{center}
\begin{minipage}[b]{.43\linewidth}
\includegraphics[width=0.8\linewidth]{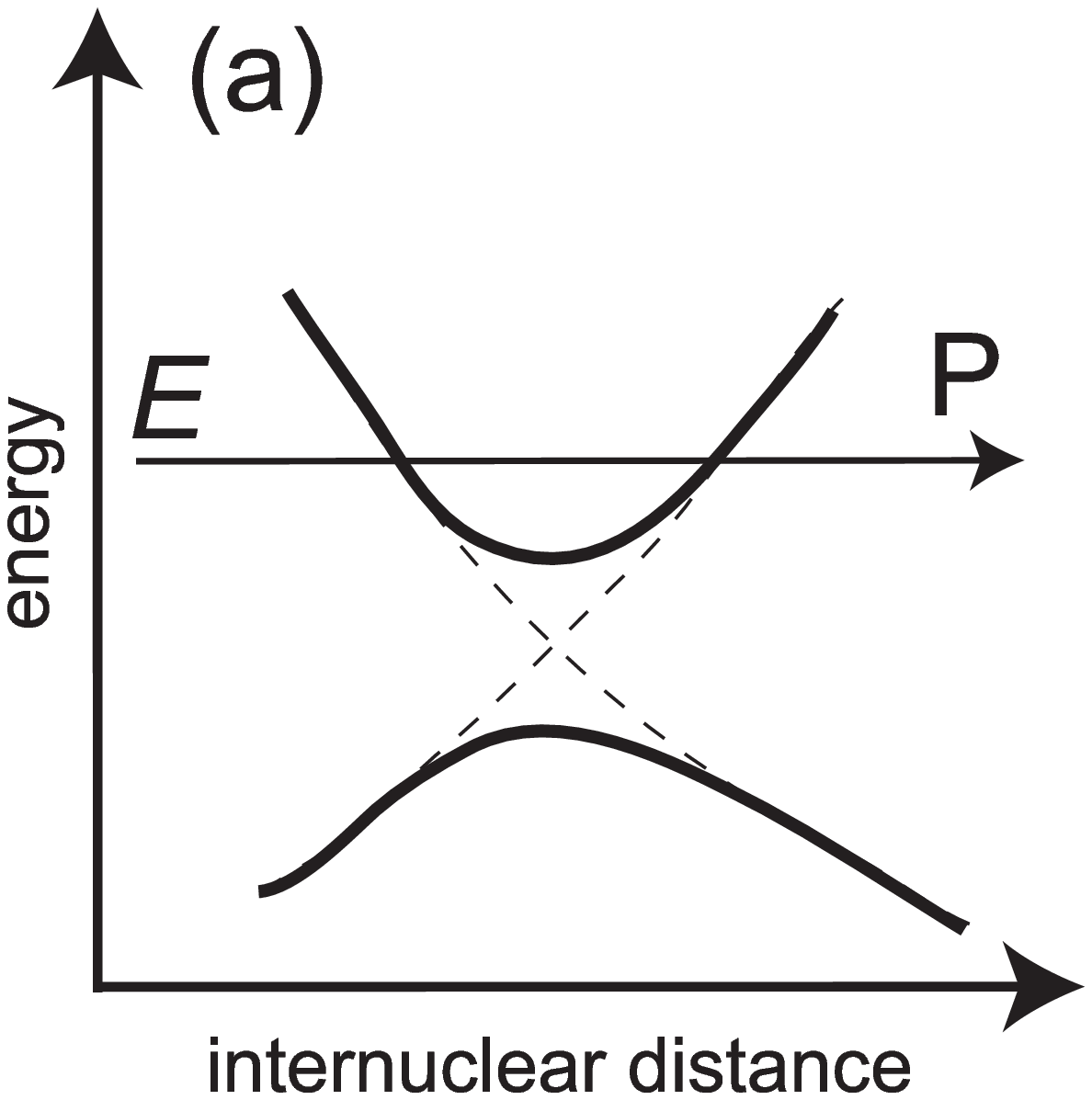}
\end{minipage}
\begin{minipage}[b]{.43\linewidth}
\includegraphics[width=0.8\linewidth]{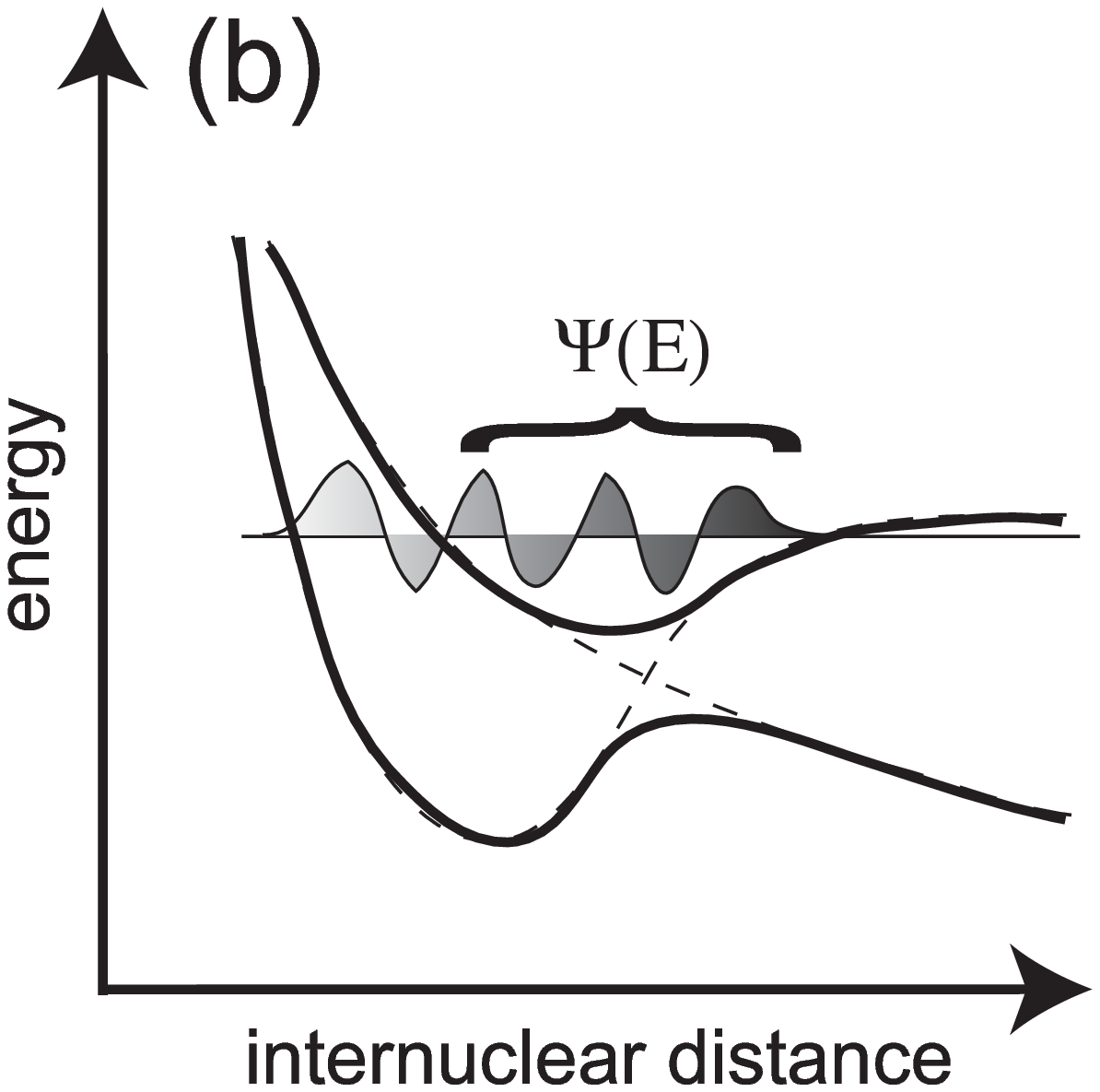}
\end{minipage}
\caption{
{\sf 
(a): Schematic picture representing 
the nonadiabatic tunneling type transition.
(b): Schematic picture representing 
the complete reflection condition in a 
diatomic molecule.
}
}
\label{fig:schematic1}
\end{center}
\end{figure}

Here we briefly summarize the relevant 
portion of the Zhu-Nakamura theory to describe the complete 
reflection phenomenon \cite{NTN00,ZTN01,Nakamura02}. 
The overall transmission probability $P$ at enery $E$ in the 
nonadiabatic tunneling type curve crossing 
[see Fig.\ \ref{fig:schematic1} (a)] is given by
\begin{equation}
P = \frac{4 \cos^2 \Psi(E)}{4 \cos^2 \Psi(E) +p^2/(1-p)}
\label{overall}
\end{equation}
with
\begin{equation}
p = \exp \left\{ 
-\frac{\pi}{4 \sqrt{\alpha \beta}} 
\sqrt{\frac{2}{1+\sqrt{1-\beta^{-2}f}}} 
\right\}
\quad
{\rm and}
\quad
\Psi(E) = \sigma-\phi_s -g,
\end{equation}
where $p$ represents the nonadiabatic transition probability
for one passage of the crossing region, and 
\begin{eqnarray}
\phi_s 
&=& \frac{\delta}{\pi} 
\ln  \left(\frac{\delta}{\pi}\right) -\frac{\delta}{\pi} 
-\arg \Gamma \left(i \frac{\delta}{\pi} \right)-\frac{\pi}{4},
\\
\sigma &=& \int_{t_1}^{t_2} \sqrt{\frac{2m[E-E_2(x)]}{\hbar}} dx,
\label{eq:int}
\\
g &=& \frac{0.23 \alpha^{1/4}}{\alpha^{1/4}+0.75} 40^{-\sigma},
\\
f &=& 0.72-0.62 \alpha^{0.715},
\\
\delta &=&
\frac{\pi}{16 \sqrt{\alpha \beta}} 
\frac{\sqrt{6+10 \sqrt{1-\beta^{-2}}}}
{1+\sqrt{1-\beta^{-2}}},
\\
\alpha &=& \frac{(1-\gamma^2) \hbar^2}{m (x_b-x_t)^2 (E_b-E_t)},
\label{eq:alpha}
\\
\beta &=& \frac{E-(E_b+E_t)/2}{(E_b-E_t)/2},
\end{eqnarray}
and 
\begin{eqnarray}
\gamma &=& \frac{E_b-E_t}{E_2 \left(\frac{x_b+x_t}{2} \right)
-E_1 \left(\frac{x_b+x_t}{2} \right)}.
\end{eqnarray}
Here $E_1(x) [E_2(x)]$ is the lower [upper] adiabatic potential, 
$x_t (x_b)$ and $E_t (E_b)$ represent the position of the top (bottom)
of the lower (upper) adiabatic potential and the corresponding energy
at $x_t (x_b)$,
$t_1$ and $t_2$ are the turning points on $E_2(x)$ at energy $E$,
and $m$ is the reduced mass of the system.
It is easily seen that the transmission does not occur 
at all when 
\begin{equation}
\Psi(E)=(n+1/2) \pi
\quad (n=0,1,2,\cdots)
\label{eq:mani}
\end{equation}
is satisfied.
This is the condition for the complete reflection
in the potential system schematically 
shown in Fig.\ \ref{fig:schematic1} (a).
This phenomenon is due to the quantum mechanical interference
effect between the wave trapped on the upper adiabatic potential
and the wave transmitting along the lower adiabatic potential 
without any transition to the upper state, and occurs 
irrespective of the potential shape and coupling strength.

In the case of a diatomic molecule as shown 
in Fig.\ \ref{fig:schematic1} (b),
predissociation cannot occur if the condition (\ref{eq:mani})
is satisfied in the region designated in the figure.
A big advantage of the present scheme is that this 
condition can always be satisfied by appropriately choosing
the CW laser frequency $\omega$, since molecular potential
curves can be shifted up and down by $\hbar \omega$ in the 
case of one-photon process.
If there are more than one dissociative states,
we may control the dissociation branching as we wish.
Thus, the key factors of this scheme are 
(1) initial preparation of an appropriate 
vibrationally excited eigenstate with energy $E_v$,
and (2) appropriate choice of 
the stationary laser frequency $\omega$.
The laser intensity coupled with the transition
dipole moment determines the diabatic coupling between
the two states and mainly controls the shape of the 
complete reflection dip.
Experimentally, (2) may be easy, but (1) might be difficult 
especially in the case of homonuclear nonpolar molecules.
Hereafter we call $\Psi_v(\omega)=\Psi(E_v+\hbar \omega)$
a complete reflection manifold.

\section{HI molecule in a laser field}

The present control scheme is applied to 
the photodissociation of an HI molecule 
in the energy range $\hbar \omega=$ 3 $\sim$ 6 eV, 
where 
three electronically excited states, $^1 \Pi_1$, $^3 \Pi_{0+}$, 
and $^3 \Pi_1$, are involved \cite{ALKB00}. 
The electronically ground state is $^1 \Sigma$ and 
is coupled to these three excited states 
through the transition dipole moment 
(see Eq.\ (\ref{eq:pot}) below).
The couplings among the latter due to the laser field 
can be neglected, 
because the corresponding transitions are off-resonant.
The coupling between $^1 \Pi_1$ and $^3 \Pi_{0+}$
induced by the spin-rotation interaction is also neglected 
as in \cite{BAB01}, since it is very weak.

Thus, we have the following 4 $\times$ 4 Schr\"odinger equations: 
\begin{equation}
i \hbar \frac{\partial}{\partial t} 
\phi(R,t)= {\cal H}(t) \phi(R,t)
=
\left( -\frac{\hbar^2}{2 m} \frac{d^2}{dR^2}+ {\cal V}(t) \right) 
\phi(R,t),
\label{eq:Sch}
\end{equation}
where 
\begin{equation}
{\cal V}(t)= 
\left(
\begin{array}{cccc}
V_1(R) & -\mu_{12} (R)F(t) & -\mu_{13} (R)F(t) & -\mu_{14} (R)F(t) 
\\
-\mu_{12} (R)F(t) & V_2(R) & 0 & 0
\\
-\mu_{13} (R)F(t) & 0 & V_3(R) & 0
\\
-\mu_{14} (R)F(t) & 0 & 0 & V_4(R)
\end{array}
\right).
\label{eq:pot}
\end{equation}
Here $i=1,2,3,4$ corresponds to the  
$^1 \Sigma$, $^1 \Pi_1$, $^3 \Pi_{0+}$, 
$^3 \Pi_1$ electronic states, respectively.
$V_i(R)$ are the potential energy curves (PEC),  
$\mu_{ij}(R)$ the transition dipole coupling between 
PEC $i$ and $j$, and 
$m=126.9/127.9 \simeq 0.99$ amu is the reduced mass of HI.

\begin{figure}[htbp]
\hfill
\begin{center}
\begin{minipage}[b]{.43\linewidth}
\includegraphics[width=1.0\linewidth]{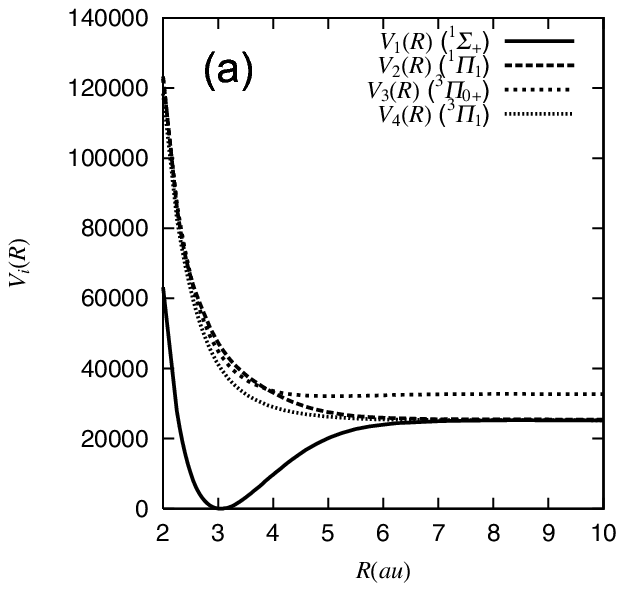}
\end{minipage}
\begin{minipage}[b]{.43\linewidth}
\includegraphics[width=1.0\linewidth]{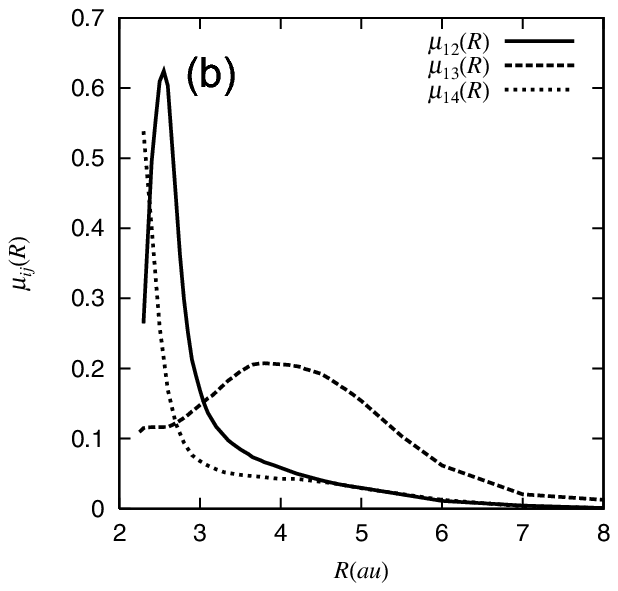}
\end{minipage}
\caption{
{\sf 
(a): Ab initio PECs for HI.
Unit for the vertical axis is cm$^{-1}$.
(b): Ab initio transition dipole moments between the electronically 
ground and excited states. Taken from \cite{Balakrishnan}.
Unit for the vertical axis is au.
}
}
\label{fig:pes}
\end{center}
\end{figure}

Ab initio data for $V_i(R)$ and $\mu_{1j}(R)$ are shown 
in Fig.\ \ref{fig:pes}. 
These data \cite{Balakrishnan} 
are based on the ab initio calculations by 
Alekseyev {\it et al}. \cite{ALKB00}.
In Refs.\ \cite{LS88,KS93}, the authors 
used a Morse potential for the ground state 
and exponential functions for the excited states,
but this approximation has turned out to be poor.
It should be noted that the dipole moments between the 
electronically ground and excited states  
significantly depend on the inter-nuclear distance 
[see Fig.\ \ref{fig:pes} (b)]; and 
the Condon approximation must be very poor. 
Note also that the two electronic states 
$^1 \Pi_1$ and $^3 \Pi_1$ 
are asymptotically connected to the same 
channel: HI $\rightarrow$ H+I.
This makes the control of branching rather 
complicated, because {\it both} states should 
be stopped {\it simultaneously}.
In the following numerical calculations,
we use a laser field
\begin{equation}
F(t)=F_0 \cos (\omega t) \Theta(t),
\end{equation}
where $\Theta(t)$ represents an envelope function which 
should be wide and smooth enough. 
This is used simply to avoid any unnecessary transitions 
due to the sudden switching of the field.
In the present calculation, we have used 
$F_0=2.8 \times 10^{9}$ V/m 
which corresponds to $I_0= 
\varepsilon_0 c F_0^2/2 \simeq 1$ TW/cm$^2$. 
This is a moderate intensity which does not 
cause any strong multiphoton transitions.

We have used the 1D spline fitting to the PECs and 
dipole moments as done by Balakrishnan {\it et al}. \cite{BAB01}.
From the theory described in the previous section,
we construct the complete reflection manifolds:
$\Psi_v^{(i)}(\omega)$
for each electronically excited state $i$.
Since the dipole moment is not very strong,
i.e., less than 0.2 a.u. in the crossing region,
$\phi_s \simeq \pi/4$ and $\delta \ll 1$ 
for the laser intensity less than $\sim$ 10 TW/cm$^2$,
which is close to the diabatic limit. 

The complete reflection phenomenon itself can be analyzed 
with use of the time independent picture in the 
Floquet (or dressed state) representation. 
The relevant time-independent Schr\"odinger 
equation is 
\begin{equation}
\left( -\frac{\hbar^2}{2 m} \frac{d^2}{dR^2}+ {\cal W}(R) \right) 
\phi(R)=E \phi(R),
\label{eq:Sch2}
\end{equation}
where 
\begin{equation}
{\cal W}(R)= 
\left(
\begin{array}{cccc}
V_1(R)+\hbar \omega & -\mu_{12} (R)F_0/2 & -\mu_{13} (R)F_0/2 
& -\mu_{14} (R)F_0/2 
\\
-\mu_{12} (R)F_0/2 & V_2(R) & 0 & 0
\\
-\mu_{13} (R)F_0/2 & 0 & V_3(R) & 0
\\
-\mu_{14} (R)F_0/2 & 0 & 0 & V_4(R)
\end{array}
\right).
\label{eq:pot2}
\end{equation}
The adiabatic potential $E_1(x)$ [$E_2(x)$] in section 2 
corresponds to the ground adiabatic state [one of the 
excited adiabatic state] obtained by diagonalizing ${\cal W}(R)$.

\begin{figure}[htbp]
\hfill
\begin{center}
\begin{minipage}[b]{.43\linewidth}
\includegraphics[width=1.0\linewidth]{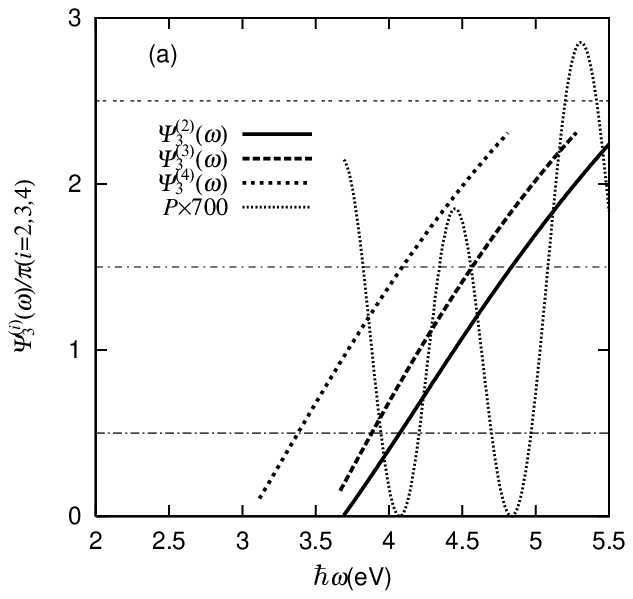}
\end{minipage}
\begin{minipage}[b]{.43\linewidth}
\includegraphics[width=1.0\linewidth]{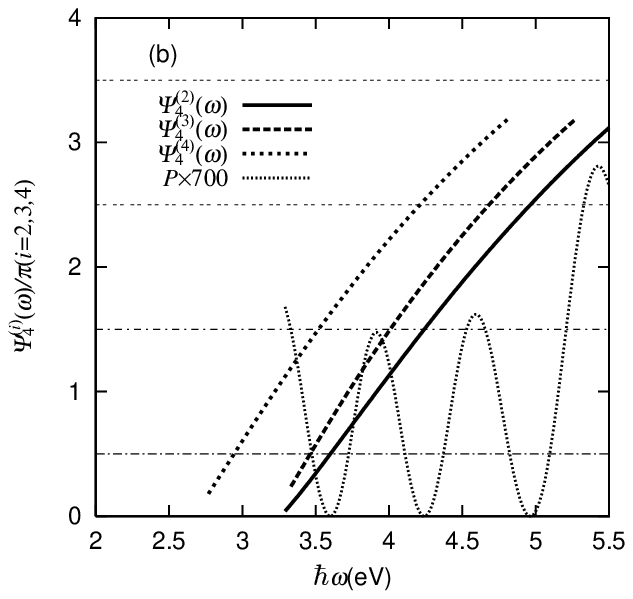}
\end{minipage}
\begin{minipage}[b]{.43\linewidth}
\includegraphics[width=1.0\linewidth]{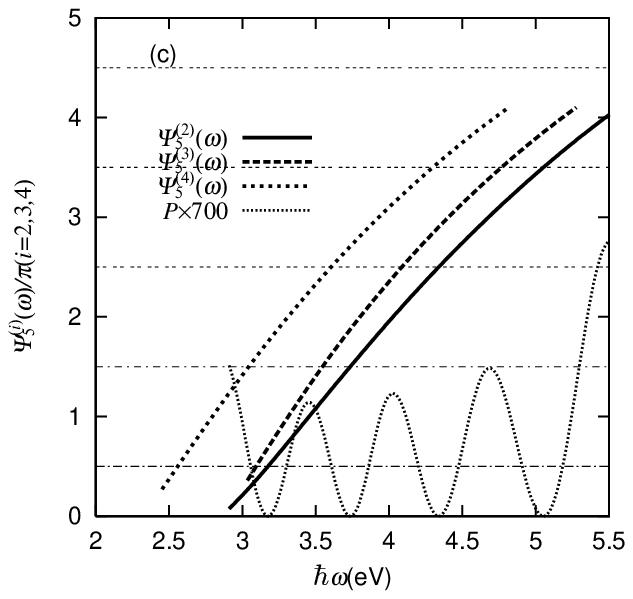}
\end{minipage}
\begin{minipage}[b]{.43\linewidth}
\includegraphics[width=1.0\linewidth]{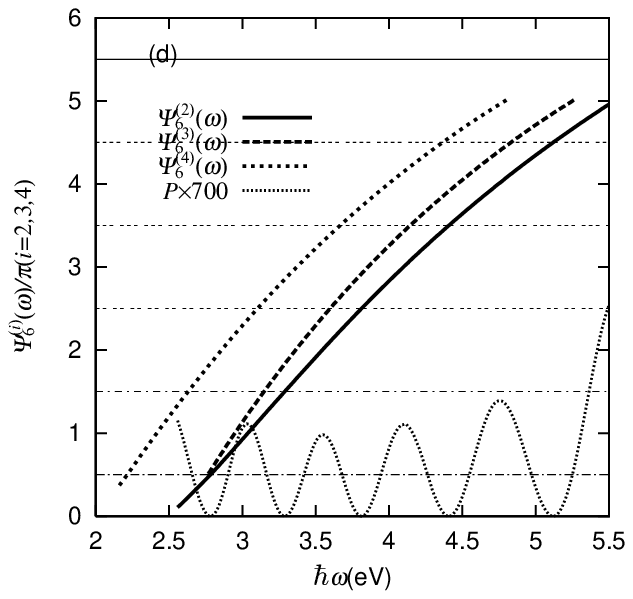}
\end{minipage}
\caption{
{\sf 
The complete reflection manifolds and 
the transmission probabilities $P$ in the 
case of the $^1 \Pi_1$ state (i=2) for (a) $v=3$,
(b) $v=4$, (c) $v=5$, and (d) $v=6$.
}
}
\label{fig:semi_manifold}
\end{center}
\end{figure}

Figures \ref{fig:semi_manifold}, 
show some manifolds for various $v$ ($v=3,4,5,6$). 
Since the vertical axis represents $\Psi_v^{(i)}(\omega)/\pi$,
the crossing points with the horizontal lines $(n+1/2)$ $(n=0,1,2, \cdots)$ 
represent the laser frequencies where the complete reflection occurs.
The transmission probabilities $P$ given by Eq.\ (\ref{overall}) 
are also shown in Fig.\ \ref{fig:semi_manifold} 
for the case of $^1 \Pi_1$ state ($i=2$). 
To make the control efficient, 
we should be able to stop the dissociation along the two states 
$^1 \Pi_1$ and $^3 \Pi_1$ simultaneously, as mentioned above.
As seen from Fig.\ \ref{fig:semi_manifold} (a),
$\hbar \omega \simeq 4.1$ eV for $v=3$ nicely
satisfies this condition. 
However, this cannot be a good candidate, unfortunately,
because the energy $\simeq 4.1$ eV is a bit too high and 
the flux from the $v=0$ component degrades 
the control efficiency significantly, as will be explained later.
To circumvent such a situation, we need to 
search for such cases that require photon energies less than 4 eV.
Actually such cases can be found for $v=4$ and $v=5$ 
in the energy range $\hbar \omega \simeq  3.5 \sim 3.7$ eV
[see Fig.\ \ref{fig:semi_manifold} (b) and (c)].
In these cases, the photodissociation branching 
is dominated by HI $\rightarrow$ H+I$^*$.
On the other hand,
we can easily find such a laser frequency that 
completely blocks the dissociation along the $^3 \Pi_{0+}$ state. 
In the next section, we carry out the 
full quantum wave packet dynamics and 
confirm the above prediction by the semiclassical theory.

\section{Wavepacket dynamics calculations}

In order to prepare a vibrationally excited state 
in the electronically ground state, 
we have employed the screening method \cite{TH95}.
The actual calculation of wavepacket dynamics 
is carried out by utilizing the sixth-order symplectic integrator 
method \cite{TI93} with use of the orthogonal 
transformation $U$ which diagonalizes ${\cal V}(t)$, i.e., 
\begin{eqnarray}
\phi(t+\Delta t) &=&
e^{-i c_8 K \Delta t} e^{-id_7 {\cal V}(t) \Delta t} 
e^{-i c_7 K \Delta t} e^{-id_6 {\cal V}(t) \Delta t} 
\nonumber
\\
&&
\times 
\cdots
e^{-id_1 {\cal V}(t) \Delta t}   
e^{-i c_1 K \Delta t} 
\phi(t) + {\cal O}(\Delta t^7)
\nonumber
\\
&=& 
e^{-i c_8 K \Delta t} U (U^t e^{-id_7 {\cal V}(t) \Delta t} U) U^t
e^{-i c_7 K \Delta t} U (U^t e^{-id_6 {\cal V}(t) \Delta t} U) U^t
\nonumber
\\
&&
\times 
\cdots
U (U^t e^{-id_1 {\cal V}(t) \Delta t} U) U^t   
e^{-i c_1 K \Delta t} 
\phi(t) + {\cal O}(\Delta t^7)
\end{eqnarray}
where $K=-\frac{\hbar^2}{2 m} \frac{d^2}{dR^2}$ 
is the kinetic energy operator, and $c_i$ and $d_j$ are the 
coefficients in the symplectic method \cite{TI93}.
We have used the following parameters: 
$\Delta R=7.8 \times 10^{-3}$ a.u. and $\Delta t= 0.043$ fs.
This method is faster than the usually employed 
split-operator method.

To avoid artificial reflection at the rightend of the PECs, 
we put an imaginary absorbing potential 
in the asymptotic region $R =9 \sim 10$ a.u.
of each electronically excited state \cite{NB89,Zhang99}. 
We calculate the time-integrated fluxes $J_i(t)$ 
on PEC $i$ according to 
\begin{equation}
J_i(t)=
\left.
\int_0^{t} dt'
\frac{\hbar}{m} {\rm Im} 
\left \{ 
\phi^*_i(R,t') \frac{d}{dR} \phi_i(R,t')
\right \}
\right|_{R=R_c}, 
\end{equation}
where 
$\phi_i(R,t)$ is the wavefunction on the surface $i$, and 
$R_c$ is an asymptotic position taken here to be 6 a.u..
Actual calculations of the fluxes 
are carried out by using the five-point numerical differentiation \cite{KM90}.
The total duration time to calculate the time-integrated fluxes is $T=$ 3.5 ps.

\begin{figure}[htbp]
\hfill
\begin{center}
\begin{minipage}[b]{.43\linewidth}
\includegraphics[width=1.0\linewidth]{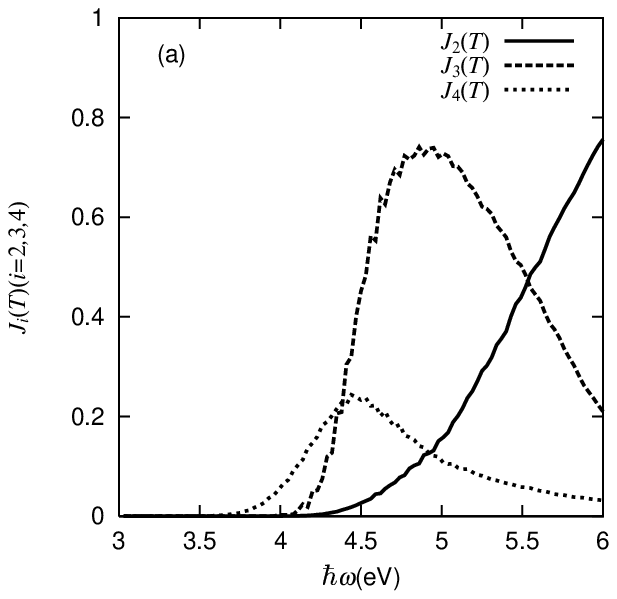}
\end{minipage}
\begin{minipage}[b]{.43\linewidth}
\includegraphics[width=1.0\linewidth]{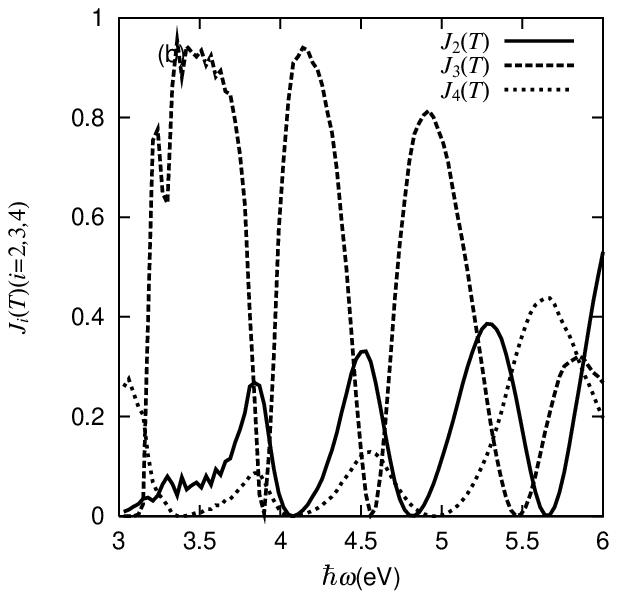}
\end{minipage}
\begin{minipage}[b]{.43\linewidth}
\includegraphics[width=1.0\linewidth]{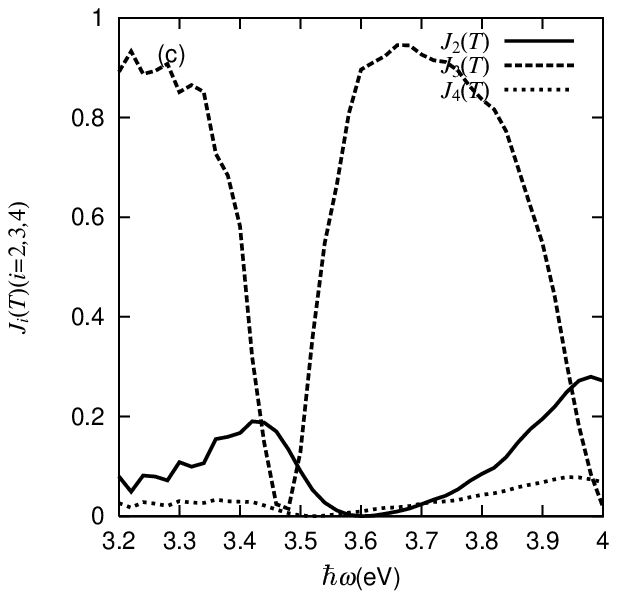}
\end{minipage}
\begin{minipage}[b]{.43\linewidth}
\includegraphics[width=1.0\linewidth]{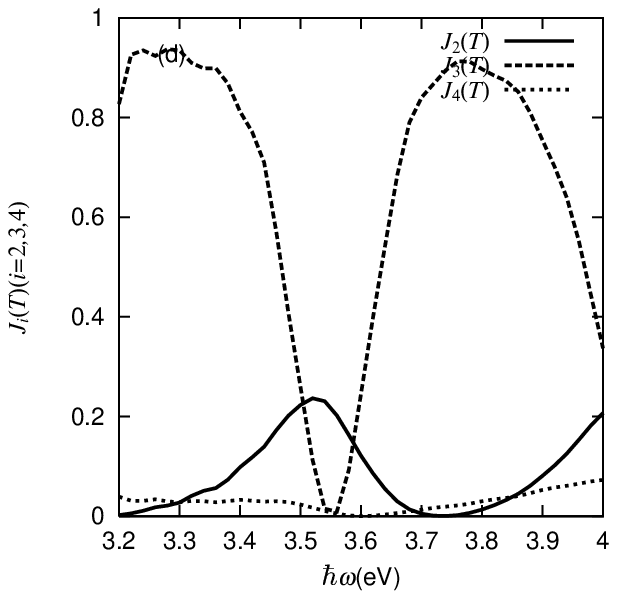}
\end{minipage}
\caption{
{\sf 
The time-integrated fluxes at $T=3.5$ ps as a function of $\hbar \omega$ 
for (a) $v=0$, (b) $v=3$, (c) $v=4$, and (d) $v=5$.
}
}
\label{fig:flux_end}
\end{center}
\end{figure}

\begin{figure}[htbp]
\hfill
\begin{center}
\begin{minipage}[b]{.43\linewidth}
\includegraphics[width=1.0\linewidth]{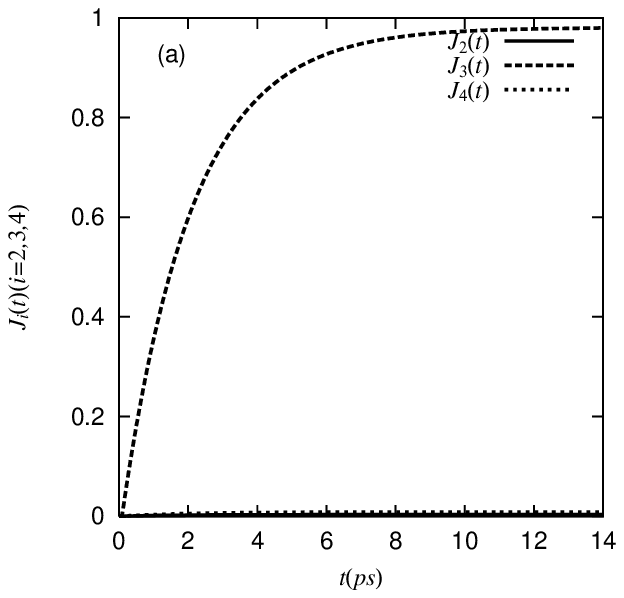}
\end{minipage}
\begin{minipage}[b]{.43\linewidth}
\includegraphics[width=1.0\linewidth]{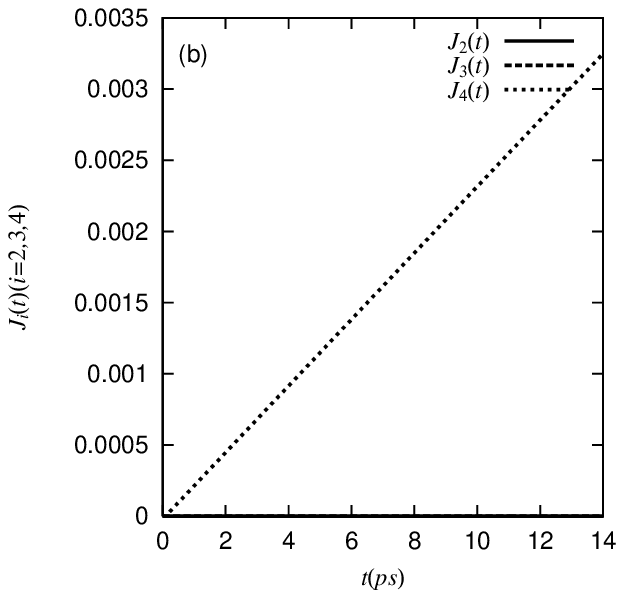}
\end{minipage}
\caption{
{\sf 
Time variation of the time-integrated fluxes $J_i(t)$ for 
(a) $v=4$ and (b) $v=0$ at $\hbar \omega \simeq 3.58$ eV.
}
}
\label{fig:flux_time}
\end{center}
\end{figure}

Figures \ref{fig:flux_end} show the 
time-integrated fluxes at the final time $T=3.5$ ps as a function of the 
laser energy $\hbar \omega$.
The time-integrated fluxes $J_i(t)$ represents 
the probability of the state $i$ at $t=T$, and 
is equal to the overall dissociation probability at $t=\infty$.
The initial excited vibrational state is prepared at $t=0$.
The final time $T=3.5$ ps is chosen rather arbitrarily
in order to save the CPU time (see Fig.\ \ref{fig:flux_time}).

Figure \ref{fig:flux_end} (a)
is the result when the initial state is $v=0$.
From this figure we can easily see (1) $v=0$ is 
not suitable for the control of photodissociation 
branching of HI, and (2) the $v=0$ component 
degrades the control efficiency, 
if $\hbar \omega$ is higher than $\simeq 4$ eV.
Figure \ref{fig:flux_end} (b) shows 
the result for $v=3$, in which, 
as predicted in Fig.\ \ref{fig:semi_manifold} (a),
the time-integrated fluxes on the two states 
$^1 \Pi_1$ and $^3 \Pi_1$ vanish at $\hbar \omega \simeq 4.1$ eV.
As mentioned above, however, 
the flux from $v=0$ at this energy ($\simeq 4.1$ eV) 
is not negligible and the control efficiency may be very much
deteriorated, since the $v=0$ component is considered 
to be dominant in the actual experimental condition unless 
the scheme of complete excitation is employed to 
prepare the initial $v=3$ state \cite{NTN02a,NTN02b}.
In order to avoid this circumstance, we can choose 
$v=4$ at 
$\hbar \omega \simeq 3.58$ eV [see Fig.\ \ref{fig:flux_end} (c)]
or 
$v=5$ at 
$\hbar \omega \simeq 3.68$ eV [see Fig.\ \ref{fig:flux_end} (d)].
To confirm this, in Figs.\ \ref{fig:flux_time}, 
we show the time variation of the 
time-integrated fluxes at $\hbar \omega=3.58$ eV 
for $v=4$ [Fig.\ \ref{fig:flux_time} (a)] and 
$v=0$ [Fig.\ \ref{fig:flux_time} (b)].
The undesirable fluxes along the states $^1 \Pi_1$ and $^3 \Pi_1$ 
from $v=4$ and from $v=0$ component of the ground 
electronic state are negligibly small.
The inverse case that  HI $\rightarrow$ H+I is dominant
can be easily achieved by choosing the laser frequency 
at $\hbar \omega \simeq 3.47$ eV with $v=4$, for instance 
[see Fig.\ \ref{fig:flux_end} (c)].
Even the initial vibrational state lower than $v=4$ 
can be used, if we want to stop only one channel correlated 
to I$^*$.

\section{Concluding remarks}

Controlling photodissociation branching between
HI $\rightarrow$ H + I and HI $\rightarrow$ H + I$^*$
was discussed from the viewpoint of the complete reflection
phenomenon and analyzed by the semiclassical Zhu-Nakamura
theory of nonadiabatic transition.
The nearly complete control was shown to be achieved 
despite the fact that the three electronically excited 
states are involved,
if the initial vibrational state and the CW laser 
frequency are appropriately selected.
This was also confirmed by carrying out much more time consuming 
quantum wavepacket calculations.
The enhancement of the branching ratio by using 
vibrationally excited states has been reported by 
Alekseyev {\it et al.} \cite{ALKB00} and by
Kalyanaraman and Sathyamurthy \cite{KS93}.
Using their own accurate ab initio potential curves, 
the same ones as those used here, the former authors 
suggested to use $v=1$ and $v=2$ at relatively high laser 
frequencies. 
The latter authors employed the inaccurate potential curves 
of Levy and Shapiro \cite{LS88} and reported the best 
branching of I$^*$ formation with $v=4$ again at 
relatively high energies. In both of these works the population
of the initial vibrationally excited state was assumed to be 
hundred percents, and the possibly large contribution from
$v=0$ was neglected. Besides, any clear and precise picture
such as that based on the complete reflection phenomenon 
as done in this work has not been provided.

In the present treatment, the rotational degree of freedom 
has been disregarded as in Refs.\ \cite{ALKB00,BAB01,KS93}.
In order to compare with any real experiment, it is required 
to take into account the effects of initial rotational 
state distribution, depending on the experimental condition.
The completeness would be deteriorated by 
the distribution to some extent, but the control may
be achieved to good extent.
This will be discussed elsewhere in near future.
The present idea can be applied to 
other diatomic molecules, and even to triatomic molecules,
if appropriate conditions are satisfied. 
Applications to Cl$_2$ \cite{SSKM01}, 
HOD, and N$_2$O are planned.

We thank Dr.\ N.\ Balakrishnan for providing us 
ab initio data of HI, and Dr.\ K.\ Nagaya 
for pointing us the importance of the initial state preparation. 
Acknowledgements
are also due to C.\ Zhu, H.\ Kamisaka, 
T.\ Yasuike, S.\ Nanbu, H.\ Katayanagi, and K.\ Hoki.
This work was partially supported by the research grant No.\ 
10440179 
from the Ministry of Education, Culture, Sports,
Science, and Technology of Japan.

\end{document}